\documentstyle[prd,aps,psfig,floats,epsfig]{revtex}

\def\beq{\begin{equation}} 
\def\eeq{\end{equation}} 
\def\bea{\begin{eqnarray}} 
\def\eea{\end{eqnarray}}

\def\flphi2{\langle\delta\phi^2\rangle} 
\def\flchi2{\langle\delta\chi^2\rangle}

\begin{document} 
\bibliographystyle{prsty} 
\twocolumn[\hsize\textwidth\columnwidth\hsize\csname 
@twocolumnfalse\endcsname 
\title{Exponential Growth of Particle Number far from the Parametric Resonance Regime} 
\author{Fernando da Rocha Vaz Bandeira de Melo}
\address{Instituto de F\'{\i}sica Gleb Wataghin, Univesidade Estadual de
Campinas, Campinas, SP, Brazil}
\author{Robert H. Brandenberger}
\address{Physics Department, Brown University, Providence, RI, 02912, USA \\ 
and \\
Theory Division, CERN, CH-1211 Gen\`eve 23, Switzerland}
\author{Adolfo Maia Junior} 
\address{Instituto de Matem\'atica, Estat\'{\i}stica e Computa\c c\~ao
Cient\'{\i}fica, Universidade Estadual de Campinas, Campinas, SP, Brazil}
\date{\today} 
\maketitle

\begin{abstract} 
 
Parametric resonance has received a considerable amount of interest as a good 
mathematical model to describe the initial stages of the reheating phase 
(matter creation) in inflationary cosmology. It is also known that 
exponential particle creation can occur in situations which do not fall
in the parametric resonance regime characterized by oscillations 
of the inflaton field about its minimum. Here we present a new
analytical approach to exponential particle production which can
occur when the inflaton is far from the minimum of its potential. Crucial
for this effect is a term in the equation of motion which acts like
a negative mass square term, as occurs for tachyonic preheating and 
negative coupling
particle production. Our techniques apply in models with a
strong coupling between matter fields $\chi$ and the inflaton $\phi$, 
or in some models in which the inflaton has a large amplitude of oscillation. 
Note that our analysis yields results which are quite model dependent. 
Exponential growth occurs in a model with interaction Lagrangian 
$-g M_{pl}\phi\chi^2$. However, for the interaction Lagrangian 
$-g^2\phi^2\chi^2$, our formalism shows that in the large coupling 
limit there can only be exponential particle production when $\phi$ crosses 0. 
 
\end{abstract} 
 
\vspace{0.35cm} 
] 
 

\section{Introduction}
\label{intro}

Inflationary Cosmology is currently the most widely accepted model
of the very early Universe. Inflation is expected to end in a
period in which the inflaton field $\phi$ is evolving rapidly
and, finally, oscillating about the ground state of its potential.
During this phase - commonly called the reheating phase -, energy is 
transferred from the inflaton field to ordinary matter, here
modelled by another scalar field $\chi$, via processes which depend 
on the coupling between $\phi$ and $\chi$ in the interaction Lagrangian.
The details of this energy transfer are important for several key
issues in cosmology, among them baryogenesis, the relic gravitino
abundance, and the possible production of topological defects.
For example, in order for the GUT-scale baryogenesis mechanism to
be effective, the temperature of matter after reheating has to be
comparable or higher than the GUT scale $T = 10^{16} GeV$ (for recent
reviews see e.g. \cite{Dolgov:1992fr}).
If the reheating process is slow (i.e. the decay rate of the inflaton is
small compared to the Hubble expansion rate at the end of inflation), as
is predicted using a perturbative analysis of inflaton decay 
\cite{Dolgov:1982th,Abbott:1982hn}, 
then the temperature after reheating is predicted to
be much lower than the GUT scale.
  
Recently, however, it was realized \cite{Traschen:1990sw} 
that in many models of inflation the energy transfer takes place much 
more rapidly, based on the phenomenon
of {\it parametric resonance}. The basic idea is the following:
while the inflaton field is oscillating about its minimum, it induces a 
periodically varying time dependence in the mass of the matter field, 
which in turn leads to exponential growth of the matter fields which 
corresponds to an explosive creation of particles. The theory was
put on a firm mathematical basis in \cite{Kofman:1994rk,Shtanov:1995ce} (these
analyses included the expansion of the Universe), and a comprehensive and
detailed study was given in \cite{Kofman:1997yn}.    
According to the new picture, 
the whole reheating process is divided into two phases. The first one is 
the phase of parametric resonance, called {\it preheating} 
\cite{Kofman:1994rk}. This is an out-of-equilibrium process during 
which the growth of the number of particles is exponential but only occurs 
for momenta in certain resonance bands. In the
second phase, interactions between the matter particles and back reaction 
effects become crucial and lead to thermalization of matter 
\cite{Felder:2001hr}. 
Although parametric resonance is an attractive mechanism for particle 
production, there are other possibilities to obtain exponential
particle production, for example 
{\it instant preheating} \cite{Felder:1999vq}, 
{\it chaotic preheating} \cite{Zibin:2001cp} and 
{\it tachyonic preheating} \cite{Felder:2001hj,Felder:2001kt}. 
In instant preheating, particles are produced exponentially when the 
inflaton field $\phi$ crosses the minimum value and the interaction 
potential is negative (see also \cite{Greene:1997ge} for a discussion
of negative coupling instability), in chaotic 
preheating the exponential increase in the perturbations is driven by the 
chaotic dynamics of the background, and tachyonic preheating is particle
production driven by a negative effective mass square term which occurs during 
spinodal decomposition in models of new and hybrid 
inflation \cite{Linde:1994cn}.

In this paper we present a new mathematical approach to exponential particle 
production at the end of inflation but far from the parametric resonance
regime. Our effect is generated by a mass square term for the matter 
fields which
is negative for certain periods in the evolution of the inflaton field 
(in this respect the mechanism is exactly what happens in 
tachyonic \cite{Felder:2001hj,Felder:2001kt} and negative 
coupling \cite{Greene:1997ge} preheating). However, our analytical method
applies also for quite general nonlinear correction terms in the
equations of motion. Note that we are not assuming 
periodic variation of the external field. Typically, our mechanism operates 
for large values of $|\phi|$ but only for particular forms of the
interaction Lagrangian (in particular we need a large effective coupling 
constant). We derive our
effect making use of some results on the asymptotics of solutions of 
second order ordinary differential equations \cite{livrinho}. 

In next section
we summarize the relevant mathematical results quoted from \cite{livrinho}. In
Section \ref{aplic} we apply them to differential equations which describe the
evolution of the matter fields during reheating in the presence of a 
dynamical inflaton field. Finally, we briefly discuss the results and
point out analogies and differences with previous work.

\section{Mathematical Approach}
\label{math}

The results below are based on the asymptotic analysis of second
order differential equations as described e.g. in reference \cite{livrinho}.

We consider the ordinary differential equation
\begin{equation} \label{eq1}
\ddot y - p(t,\sigma)y=0 \; ,
\label{dif}
\end{equation}
where $p(t,\sigma)$ is a real function for values of $t$ in some interval $I$,
and for values of $\sigma$ in an interval $E$. The derivatives are with respect
to $t$. Both $t$ and $\sigma$ are real
variables, and in our applications to cosmology $t$ will be time, whereas $\sigma$
is some external parameter. In this section we recall some
results concerning the behavior of the solutions $y_1(t)$ and $y_2(t)$ of
(\ref{eq1}) for values of $t$ in $I$ when $\sigma\rightarrow+\infty$.

We assume that in the limit $\sigma\rightarrow+\infty$, the function
$p(t, \sigma)$ can be expanded in the following way:
\begin{equation}
p(t ,\sigma) \, \approx \, \sigma^r \sum^\infty_{n=0}
p_n(t)\sigma^{-\frac{r n}{2}}\; ,
 \label{expan}
\end{equation}
where $r$ is a positive rational number.
We work in subintervals $I'\subset I$ where $p_0(t)$ does not
vanish, but could have at most double poles in $I'$.

With these hypotheses, by a theorem in reference \cite{livrinho},
it can be shown that:

\noindent a) In subintervals $I'\subset I$ where $p_0(t) > 0$, and when
$\sigma\rightarrow\infty$ : 
\begin{eqnarray}
y_1(t)=&& \{p_0(t)\}^{-1/4}\exp\biggl\{\int^t
[\sigma^{r/2}(p_0(t'))^{1/2}+\nonumber\\
&&\frac{1}{2}p_1(t')(p_0(t'))^{-1/2}]dt'\biggl\}
\biggl\{1+O\biggl
(\frac{1}{\sigma}\biggl )\biggl\}
\label{we}
\end{eqnarray}
and
\begin{eqnarray}
y_2(t)=&&\{p_0(t)\}^{-1/4}\exp\biggl\{-\int^t
[\sigma^{r/2}(p_0(t'))^{1/2}+\nonumber\\
&&\frac{1}{2}p_1(t')(p_0(t'))^{-1/2}]dt'\biggl\}\biggl\{1+O\biggl
(\frac{1}{\sigma}\biggl )\biggl\}
\label{we2}
\end{eqnarray}

\noindent b) In subintervals $I'\subset I$ where $p_0(t) < 0$, and when
$\sigma\rightarrow\infty$ :
\begin{eqnarray}
y_1(t)=&&c\{-p_0(t)\}^{-1/4}\sin\biggl\{\int^t
[\sigma^{r/2}(-p_0(t'))^{1/2}+\nonumber\\
&&\frac{1}{2}p_1(t')(-p_0(t'))^{-1/2}+a]dt'\biggl\}\biggl\{1+O\biggl
(\frac{1}{\sigma}\biggl )\biggl\}
\label{wosc}
\end{eqnarray}
and;
\begin{eqnarray}
y_2(t)=&&c\{-p_0(t)\}^{-1/4}\cos\biggl\{\int^t
[\sigma^{r/2}(-p_0(t'))^{1/2}+\nonumber\\
&&\frac{1}{2}p_1(t')(-p_0(t'))^{-1/2}+a]dt'\biggl\}\biggl\{1+O\biggl
(\frac{1}{\sigma}\biggl )\biggl\} \, .
\label{wosc2}
\end{eqnarray}
Here, the phase angle $a$ is a real constant, and $c$ is a
possibly complex constant.

Thus, if $p_0(t) > 0$ on $I'$, in the asymptotic approximation $y(t)$ increases
(or decreases) exponentially. However, if $p_0(t) <0 $, we obtain
oscillating solutions. This theorem generalizes the trivial dynamics of a
harmonic oscillator with positive or negative square mass to the case where
the coefficient representing the square mass is not constant in time (but does
not cross zero, either). Thus, in the adiabatic limit in which the mass varies only
slowly, the results are easy to understand from the point of view of physics.
These results will be used in the next
section in order to study exponential growth of particle number
at the end of inflationary phase of the evolution of the Universe.

\section{Applications}
\label{aplic}

In this section we use the above theorem to get some
information on particle creation induced by the inflaton field. In
order to be as general as possible, we do not prescribe any
specific equation of motion for the inflaton field, but only
consider some general properties needed, which are shared
by several models of chaotic inflation. As indicated, in the following
$\phi$ denotes the inflaton, and $\chi$ a scalar matter field. As in
most studies of preheating, we will neglect the bare mass and 
self-interactions of the
$\chi$ field. We will also neglect the expansion of the Universe. As long
as the time scale for exponential increase obtained under this approximation
is smaller than the Hubble expansion time, we expect the approximation to
be a good one. As interaction Lagrangian we take 
$ L_{int} = - {1 \over 2} f(\phi)\chi^2$,
where the function $f$ contains a large parameter which plays the role of 
$\sigma$ in the previous section. This parameter can be a coupling constant -
it can also be the amplitude of the inflaton during a period of oscillation.

The equation of motion of the k-th Fourier mode of the scalar matter field
$\chi$ is given by :
\begin{equation}
\ddot \chi_k -(-k^2-f(\phi))\chi_k \, = \, 0\; .
 \label{mov}
\end{equation}
Comparing this with equation (\ref{dif}) we see that:
\begin{equation}
p(\phi) \, = \, -k^2 - f(\phi)\; .
 \label{pgeral}
\end{equation}


As mentioned above, we can write $f(\phi) = \sigma h(\phi)$,
where $\sigma$ is a large parameter. Then we can write the above  equation as

\begin{equation}
p(\phi) \, = \, -k^2 - \sigma h(\phi)\; .
 \label{pgeral2}
\end{equation}

Comparing with the expansion (\ref{expan}) we get

\begin{eqnarray}
&& p_0(t)\, = \, -\sigma h(\phi(t))\; ;\nonumber\\
&& p_2(t) \, = \, -k^2\; ;\\
&& p_n(t) \, = \, 0 \; ;\; n \neq 0,2\nonumber
\end{eqnarray}

Now, from the key theorem in section \ref{math} it follows that if the 
parameter $\sigma$ is large, then  
in time intervals where $h(\phi(t))$ is positive the solutions
exhibit exponential growth, whereas in those during which $h(\phi(t))$ is 
negative we get bounded oscillating solutions.

In order to connect the above results with the physics of known models we now specialize to two cases:

\subsection{Case {\bf $f(\phi)=g^2\phi^2$}}
\label{g2f2} 

In this case we can write $p(\phi)$ as
\begin{equation}
p(\phi) \, = \, -k^2 - g^2\phi_0^2\Phi^2(t)
\label{pg2f2}
\end{equation}
where $\Phi=\frac{\phi(t)}{\phi_0}$ and $\phi_0$ is the amplitude of the inflaton
field at the beginning of the oscillatory phase during which parametric resonance
occurs. Taking $\sigma =g^2\phi_0^2$ and $r=1$ in the expansion
(\ref{expan}) we have:
\begin{eqnarray}
&& p_0 \, = \, -\Phi^2(t)\; ;\nonumber\\
&& p_2 \, = \, -k^2\; ;\\
&& p_n \, = \, 0 \; ;\; n \neq 0,2\nonumber
\end{eqnarray}
Since in this case $p_0(t) < 0$ for all $t$, the solutions $\chi_k$ are
always asymptotic oscillating when $\sigma$ is big enough. Thus, in
this model no particle production occurs during time intervals when $\phi \neq 0$. 
This result sheds new light on the observations 
\cite{Kofman:1997yn,Felder:1999vq} that particle production is
concentrated at times when $\phi = 0$.

\subsection{Case $f(\phi)=g M_{pl}\phi$}
\label{g2f}

In this case:
\begin{equation}
p(\phi)=-k^2-g M_{pl}\phi(t)\; .
 \label{pg2f}
\end{equation}
We assume that the potential can be well approximated as quadratic near the
origin, and as linear for large values of $|\phi|$ ($|\phi| > M_{pl}$). The
second assumption is made in order to obtain slow-rolling of $\phi$ for
large values of $|\phi|$. In an improved analysis the slow-rolling phase
would be generated by the expansion of the Universe and the resulting
Hubble damping term in the scalar field equation of motion. Since
$\phi_0$ is of the order $M_{pl}$,
the condition to apply the theorem of Section (\ref{math}) 
(to obtain large $\sigma$) is
$k^2, m^2 \ll g M_{pl} \phi_0$, where $m$ is the inflaton mass. Thus, in particular,              
\begin{equation} \label{large}
g \, \gg \, \biggl ( \frac{m}{M_{pl}} \biggl )^2
\end{equation}

For a potential of the form $V(\phi)= m^2\phi^2 + \lambda\phi^4 $
the COBE constraints on the fluctuations generated during inflation give 
$m / M_{pl} < 10^{-6}$ and $\lambda < 10^{-12}$. Thus, (\ref{large}) becomes $ g \gg 10^{-12}$.
On the other hand, by naturalness, $ g \leq \sqrt{\lambda}$. Therefore, the large coupling
asymptotic analysis is valid in a large interesting region of values of g from $10^{-12}$ to $10^{-6}$.

\begin{figure}[htf]
\centerline{\epsfig{figure=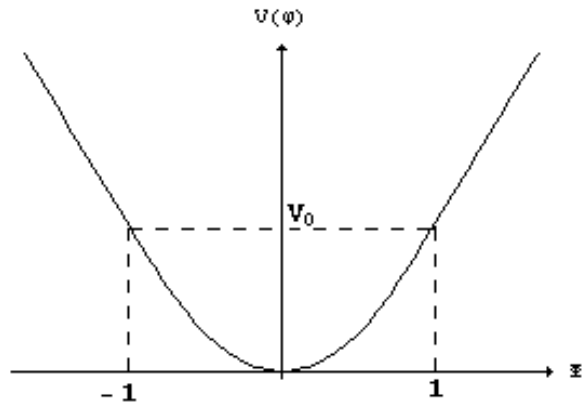}}
\caption{Sketch of a chaotic inflation potential $V(\phi)$. 
$|\Phi| > 1 $ is the slow rolling phase in which a linear approximation to $V(\phi)$ is used. The region $|\Phi|<1$ is the
region of oscillation with a quadratic approximation.}
\label{potentialfig}
\end{figure}

Taking $\sigma = g M_{pl} \phi_0$ and $r=1$ in the
expansion (\ref{expan}), and using the same variable $\Phi$ introduced in the
previous subsection, we have from (\ref{pg2f}):
\begin{eqnarray}
&& p_0(t) \, = -\Phi(t) \nonumber \\
&& p_2 \, = \, -k^2 \\
&& p_n \, = \, 0 \; ; n \neq 0,2 \, .\nonumber 
\end{eqnarray}
It follows immediately that for $\Phi < 0$ we have $p_0(t) > 0$, and by the
theorem presented in Section \ref{math} we expect exponential
particle production. 

In the following we assume that $\phi$ starts out at large negative values
in the slow-rolling phase. Once it reaches $|\Phi| \sim 1$, $\phi$ will proceed to oscillate about the minimum of the potential. Note that exponential
increase in $\chi$ is possible in both regions, as long as $\phi < 0$. In the following we analyse the two phases separately. In both phases, the equation of motion is
\begin{equation}
\frac{d^2\phi}{dt^2}+\frac{\partial V(\phi)}{\partial \phi} \, = \, 0\; .
\label{potential}
\end{equation}

\subsubsection{{\bf Slow rolling region}}
\label{roll-over}

In the slow rolling phase we take a linear approximation to the
potential, i.e., $V(\phi) = A\phi + B$, where $A$ and $B$ are constants. In the
region $\phi < - \phi_0$, the constant $A$ is negative, and
$B$ is chosen such that the potential is continuous at the transition points 
$|\Phi| = 1$.
 
Solving Equation (\ref{potential}) in this approximation gives:
\begin{equation}
\phi (t)=\frac{|A|}{2} t^2 + C t + D
\label{philinear}
\end{equation}
where $C$ and $D < 0$ are constants. Thus,
\begin{equation}
p(\phi ) = - k^2 - g^2\frac{|A|}{2} t^2 - g^2 C t - g^2 D\; .
\label{plinear}
\end{equation}

Taking $\sigma = g^2 |A|/2$ and $r=1$ in the
expansion (\ref{expan}), we obtain from (\ref{plinear}):
\begin{eqnarray}
&& p_0(t)=-(t^2 + \overline{C} t + \overline{D})\; ;\nonumber\\
&& p_2 = -k^2\; ;\\
&& p_n = 0 \; ;\; n \neq 0,2\nonumber
\end{eqnarray}
where $\overline{C}=2 C/|A|$ and $\overline{D} = 2 D/|A|$.

There are two roots $t_-$ and $t_+$ of the equation $p_0 = 0$:
\begin{equation}
t_\pm = \frac{-\overline{C} \pm \sqrt{\overline{C}^2 - 4
\overline{D}}}{2}\; .
 \label{tpm}
\end{equation}
Since $D < 0$, the roots are real. Thus, there is a time interval
$[t_-,t_+]$ during which the
solution for $\chi_k$ is asymptotically exponential. 
In fact it is easy to see that $\overline{C}\leq 0$ and
$\Delta t = t_+ - t_- \geq |\overline{C}| = - \overline{C}$.
From this inequality we conclude that the smaller $A$ or the bigger $C$,
the bigger is the time interval during which the solution has exponential 
growth. The case of $A$ small suggests that chaotic inflation, with its long 
slow roll-over section of the potential, is the most effective for particle 
production, when the system is far from the parametric resonance regime..

Note that in the slow rolling region with $\phi > \phi_0$, $p_0(t)$ is
negative and, as expected, we obtain asymptotically oscillating solutions.

\subsubsection{{\bf Oscillatory region}}
\label{minimo}

For $|\Phi| < 1$ we take a quadratic
approximation, that is, $V(\phi)=A' \phi^2$, where $A'>0$. Solving the differential equation (\ref{potential}) for $\phi$ we
get $\phi(t) = \phi_0 \cos (w t)$, where $\phi_0$ is the amplitude of oscillation and $w=\sqrt{2 A}$. Thus we have: 
\begin{equation}
p \, = \, -k^2 - gM_{pl}\phi_0\cos (w t)
\label{pcos}
\end{equation}

Taking again $\sigma = gM_{pl}\phi_0$ and $r=1$ in the expansion (\ref{expan}) we obtain from (\ref{pcos})

\begin{eqnarray}
&& p_0 \, = \, -\cos (w t)\; ;\nonumber\\
&& p_2 \, = \, -k^2\; ;\\
&& p_n \, = \, 0 \; ;\; n \neq 0,2\nonumber   \\
\label{p0minima}
\end{eqnarray}

For $\bigl( j-\frac{1}{2} \bigr)\pi<w t < \bigl( j+\frac{1}{2} \bigr)\pi$
with $j$ even, we have $p_0<0$. So if $\sigma$ is large, which can occur either if the coupling constant or the field amplitude
are large, the theorem of
Section \ref{math} give us an oscillating asymptotic solution:
\begin{equation}
\chi_k(t)=\frac{c}{\sqrt[4]{\cos (w t)}}\cos
\left\{\int^t[\sigma^{1/2}
 \sqrt{\cos (w t')} + a]dt'\right\}
\label{chiosc}
\end{equation}
where $a$ and $c$ are constants.
 
However, for time intervals with
$\bigl( j-\frac{1}{2} \bigr)\pi<w t < \bigl( j+\frac{1}{2} \bigr)\pi$ with $j$ odd, we get $p_0>0$. Then, for large $\sigma$, the asymptotic solution reads
\begin{equation}
\chi_k = \frac{1}{\sqrt[4]{-\cos (w t)}}e^{\sigma^{1/2} \int^t \sqrt{-\cos (w t')}dt'} \, ,
\label{chiexp}
\end{equation}
and in this case we get exponential production of matter.

\begin{figure}
\centerline{\epsfig{figure=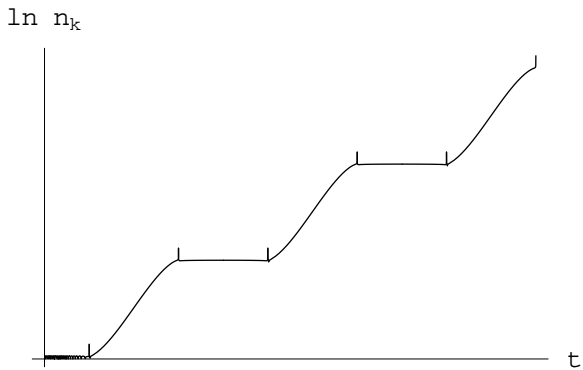}}
\caption{Pulsed Particle Creation}
\label{escada}
\end{figure}

Combining both cases, we get the time evolution of the particle number
density $n_k$ drawn in Figure (\ref{escada}). Here, $n_k$ is the number density of particles produced in the k-th mode obtained using the standard formula 
(see e.g. \cite{Kofman:1997yn})
 \begin{equation}
n_k = w_k \biggl (\frac{|\dot \chi_k |^2}{w_k^2} +|\chi_k|^2 \biggl ) -\frac{1}{2} \, ,
\label{number}
\end{equation}
where $w_k$ is the effective frequency of $\chi_k$.
Substituting 
(\ref{chiosc}) and (\ref{chiexp}) in (\ref{number}) within their respective intervals of validity, we obtain (as a function of time) a kind of pulsed particle creation (see Fig. 2).
Note that our results in this region are similar to what was previously
obtained in the context of tachyonic preheating 
\cite{Felder:2001kt}.

\section{Discussion and Conclusion}
\label{conclu}

We have presented a new analytical approach to exponential particle production
far from the parametric resonance regime. Our analysis applies to
situations when the either the absolute
value of the inflaton field is large, or the coupling constants are 
very large. The results, however, are is quite
model-specific. One obtains exponential particle production
for the interaction Lagrangian $ L_{int} =  -g M_{pl}\phi\chi^2$, but not for
$L_{int} = -g^2\phi^2\chi^2$. Crucial to have exponential particle
production is the presence of a negative mass square term in the effective
potential for the inflaton. Note that our method is robust in the sense
that the results do not change qualitatively when adding extra terms in
the equation of motion such as explicit mass terms and self interaction
terms for the $\chi$ field, as long as the additional terms are small in
amplitude in the appropriate units.  

In the rolling phase we can have exponential particle production when the
coupling constant or the slope of the potential are large. However, the length
of the time interval where this exponential growth happens decreases when the
slope $A$ of the linear potential $V(\phi) = A\phi + B$ increases. 
Then, when $A$ is large enough to use the asymptotic approximation, we do not 
have exponential matter creation for a long time. On the other hand, if $A$
is small (yielding a large interval of exponential growth), we only can 
use the asymptotic approximation provided the coupling constant 
$g$ is big enough. In this case we can
have particle creation  for a long time interval even during the rolling 
period.
Applied to the regime of fast rolling, we are in the situation of
tachyonic preheating, and our results agree with those of 
\cite{Felder:2001hj,Felder:2001kt}. 

\acknowledgments{This work was supported in part by FAPESP, Process
99/10152/3, and by the US Department of Energy
under Contract DE-FG02-91ER40688, Task A. We are grateful to
Juan Garcia-Bellido, Lev
Kofman and Andrei Linde for comments on the draft.}


\end{document}